\documentclass[onecolumn,floatfix,superscriptaddress,showpacs,showkeys,nofootinbib,preprint]{revtex4-1}
\usepackage{epsfig}
\usepackage{amssymb,latexsym,amsmath}
\newcommand{\eq}[1]{\begin{align} #1 \end{align}}

\begin{document}

\title{Identity Method for the Determination of the Moments of Multiplicity Distributions}

\author{A.~Rustamov}
\affiliation{Institut f\"ur Kernphysik, Johann Wolfgang
 Goethe Universit\"at Frankfurt, Germany}

\author{M. I. Gorenstein}
\affiliation{Bogolyubov Institute for Theoretical Physics, Kiev,
Ukraine} \affiliation{Frankfurt Institute for Advanced Studies,
Frankfurt, Germany}

\begin{abstract}
Recently the identity method was proposed to calculate second
moments of the multiplicity distributions from event-by-event
measurements in the presence of the effects of incomplete particle
identification. In this paper the method is extended for higher
moments. The moments of smeared multiplicity distributions are
calculated using single-particle identity variables. The problem
of finding the moments of the multiplicity distributions is
reduced to solving of a system of linear equations.
\end{abstract}
\maketitle

\section{Introduction}

The study of event-by-event fluctuations of chemical
(particle-type) composition in high energy nucleus-nucleus
collisions is a helpful tool to pin-down the properties of
strongly interacting matter (see, e.g., review \cite{Koch:2008ia}
and references therein).  By measuring the fluctuations one may
observe anomalies associated with transition  between hadronic and
partonic phases. Furthermore, the QCD critical point may be
signalled by a characteristic pattern in the fluctuations
\cite{fluc3,Koch:2005vg,Koch:2005pk}.
Data on event-by-event chemical fluctuations from the CERN
SPS~\cite{Afanasev:2000fu,Alt:2008ca,Kresan:2009qs} and BNL
RHIC~\cite{Abelev:2009if} were already published, and more
systematic measurements are in progress.

A serious experimental problem of event-by-event determination of
the multiplicity of different hadron species is an incomplete
particle identification caused by finite detector resolution. 
However, one can usually determine distributions $\rho_j(x)$ of
an identification observable $x$ and obtain the average multiplicities
$\langle N_j\rangle$ for different hadron species $j$.

As an artefact of particle mis-identification measured
fluctuations may be systematically reduced or enlarged. To
overcome this problem a new experimental technique, called the
{\em identity method},  was proposed in Ref.~\cite{Ga2011} for the
analysis of events with two measured particle species.
 In Ref.~\cite{Go2011} this
method was further developed to calculate the second moments of the
multiplicity distributions of more than two particle species.
In the present study we prove that the {\em identity method} can
be generalized to determine third and {\em higher moments} of the
multiplicity distributions in events consisting of an arbitrary
number of different particle species. It was already emphasized
that measurements of the third and higher moments for the
event-by-event fluctuations is important for the search of 
critical point signals in nucleus-nucleus collisions
\cite{Step,Step1}.

The paper is organized as follows. The used notations and
definitions are introduced in Section~\ref{def}. The identity
method for the second moments of the multiplicity distribution is
reviewed in Section~\ref{second_moments}. In Section~\ref{higher}
the method is generalized to third and higher moments of the
multiplicity distribution. We consider in
Section~\ref{first_revisit} the case in which the normalization of
mass distributions are unknown, and finally in section~\ref{sim}
we test the developed formalism in simulation. Section~\ref{sum}
summarizes the paper.

\section{Notations and definitions}\label{def}
We assume that particle identification (PID) is achieved by measuring a quantity $x$\footnote{Here by $x$ we refer to any experimentally measured particle property related to its identity, e.g. specific energy loss in a detector gas.}. Since any measurement is of finite resolution,
we deal with continuous distributions of $x$ denoted
by $\rho_j (x)$  ($j=1,\ldots,k\geq$ 2, with $k$ being the number
of different particle types)  and normalized as
\eq{ \label{norm-rho-i} \int dx \,\rho_j (x) = \langle N_j
\rangle~,
}
where $\langle N_j\rangle$ is the mean multiplicity of the
particle type $j$. Note that the functions $\rho_j(x)$  are found
for the different particle species using the inclusive
distribution of $x$ for all particles in all events.
The {\it identity variables} $w_j(x)$ are defined as:
\eq{\label{wi}
w_j(x)~\equiv~\frac{\rho_j(x)}{\rho(x)}~,
}
where
\eq{\label{rho-m}
 \rho (x) \equiv \sum_{j=1}^k\rho_j (x) ~.
}
The {\it complete identification} (CI) of particles corresponds to
distributions $\rho_j (x)$, which do not overlap. This case leads to
$w_j = 0$ for all particle species $i\neq j$ and $w_j = 1$ for the
$j^{th}$ species. When the distributions $\rho_j (x)$ overlap,
$w_j(x)$ can take any real value between 0 and 1.

We further introduce the event quantities, $ W_j $, for each
particle type $j$:
 \eq{\label{Wj-def}
W_j~\equiv~\sum_{i=1}^{N(n)}w_j(x_i)~,~
}
with $j=1,\cdots,k$~, and define the event averages  as
\eq{
\langle W_p\cdot\ldots \cdot W_q \rangle ~=~ \frac{1}{N_{\rm ev}}
\sum_{n=1}^{N_{\rm ev}} W_p\cdot \ldots \cdot W_q~, \label{Wj2-av}
}
where $N_{\rm ev}$ is the number of events and
$N(n)$
%=N_1(n)+\cdots+N_k(n)$
is the total multiplicity in the $n^{th}$ event. Each experimental
event is characterized by a set of measured values of PID variable $x$
$\{x_1,x_2,\ldots,x_N\}$, for which one can calculate the identity
variables: $\{w_j(x_1),w_j(x_2),\ldots,w_j(x_N)\}$, with
$j=1,\ldots,k$. Consequently, the quantities $W_j$, $W_j^2$,
$W_{p}W_q$ etc.  are uniquely defined for each event, and their
average values (see Eq.~(\ref{Wj2-av})) can be found
experimentally by straightforward averaging over all events. In
the case of CI, one finds  $W_j=N_j$, thus, Eq.~(\ref{Wj2-av})
yields
\eq{\label{CI}
%
%\langle W_j^2\rangle~=~\langle N_j^2\rangle~,~~~~
\langle W_p\cdot \ldots \cdot W_q\rangle~=~\langle N_p\cdot \ldots \cdot N_q\rangle~.
}

The $W$-quantities are the event variables as they are calculated
for each event. By averaging different combinations of these
quantities over events, the moments of the $W$-distributions
$\langle W_p \ldots W_q\rangle$ can be easily reconstructed
directly from experimental data. However, the goal is to obtain
the moments $\langle N_p\ldots N_q\rangle$ of the multiplicity
distribution, which are unknown in the case of incomplete particle
identification.
The main idea here is to find the relation between the moments of the 
$W$-quantities calculated from the measurements and the unknown moments of the
multiplicity distribution.

\section{Second Moments}\label{second_moments}

The relation between second moments of multiplicity distributions
and corresponding second moments of $W$s has been obtained in
Ref.~\cite{Go2011}.
The quantities $\langle W_j^2\rangle$ and $\langle W_pW_q\rangle$
were found as specific linear combinations of all first and second
moments $\langle N_j\rangle$ and $\langle N_j^2\rangle$, as well
as the correlation terms $\langle N_pN_q\rangle$. As the first
moments are given by Eq.~(\ref{norm-rho-i}) the mathematical
problem  of finding of all the second moments $\langle
N_j^2\rangle$ and $\langle N_pN_q\rangle$ is then reduced to
solving a system of linear equations.
%
%
%In the next Section  we demonstrate that this approach can be
%generalized for higher moments as well.

\section{Third and higher moments}
\label{higher}
In this section we derive formulae for all pure and
mixed third moments of the unknown multiplicity distributions.
Following the procedure developed for the second moments (see the
previous section) we introduce the multiplicity
($\mathcal{P}(N_{1},...,N_{k})$) and the PID probability density
($P_{i}(x)\equiv\rho_{i}(x)/\left<N_i\right>$) functions. Using
these functions we get:

%
%\begin{verbatim}
%\begin{multline}
\eq{
&\left<W_{p}^{3}\right>=\sum_{N_{1}=0}^{N_{1}=\infty}\sum_{N_{2}=0}^{N_{2}
=\infty}...\sum_{N_{k}=0}^{N_{k}=
\infty}\mathcal{P}\left(N_{1},N_{2},...,N_{k}\right)
\int{dx_{1}^{1}P_{1}(x_{1}^{1})}...\int{dx_{N_{1}}^{1}P_{1}(x_{N_{1}}^{1})}\nonumber
}
\eq{
&\times\int{dx_{1}^{2}P_{2}(x_{1}^{2})}...\int{dx_{N_{2}}^{2}P_{2}(x_{N_{2}}^{2})}
\times...\times\int{dx_{1}^{k}P_{k}(x_{1}^{k})}...
\times \int{dx_{N_{k}}^{k}P_{k}(x_{N_{k}}^{k})}\nonumber\\
&
\times \left[w_{p}(x_{1}^{1})+...+w_{p}(x_{N_{1}}^{1})+w_{p}(x_{1}^{2})+...+w_{p}(x_{N_{2}}^{2})
+w_{p}(x_{1}^{k})+...+w_{p}(x_{N_{k}}^{k}) \right] ^{3} \nonumber\\
 &=\sum_{i=1}^{k}\left<N_{i}^{3}\right>\overline{w_{p,i}}^{3}
+ 3\sum_{1\le i < l \le
k}\left<N_{i}^{2}N_{l}\right>\left(\overline{w_{p,i}}^{2}\cdot\overline{w_{p,l}}\right)
+ 3\sum_{1\le i < l \le k}\left<N_{i}N_{l}^{2}\right>
\left(\overline{w_{p,l}}^{2}\cdot\overline{w_{p,i}}\right)\nonumber \\
&+ 6\sum_{1\le i < l < m \le
k}\left<N_{i}N_{l}N_{m}\right>W^{ppp}_{ilm}
+3\sum_{i=1}^{k}\left<N_{i}^{2}\right>\left(\overline{w_{p,i}^{2}}\cdot\overline{w_{p,i}}
- \overline{w_{p,i}}^{3}\right)
\nonumber \\
& +3\sum_{1\le i < l \le
k}\left<N_{i}N_{l}\right>\big(\overline{w_{p,i}^{2}}\cdot\overline{w_{p,l}}
%\nonumber
%\\
 + \overline{w_{p,l}^{2}}\cdot\overline{w_{p,i}} -
\overline{w_{p,i}}^{2}\cdot\overline{w_{p,l}}
\nonumber \\
& - \overline{w_{p,l}}^{2}\cdot\overline{w_{p,i}}\big)
+ \sum_{i=1}^{k}\left<N_{i}\right>\big(2\overline{w_{p,i}}^{3} +
\overline{w_{p,i}^{3}}
%
%\nonumber \\
%
 -
3\overline{w_{p,i}^{2}}\cdot\overline{w_{p,i}}\big) ~,\label{Wp3}
%\end{multline}
}
%\end{verbatim}
%
%
%\begin{multline}
\eq{\label{WpWqWr}
 &\left<W_{p}W_{q}W_{r}\right> =
\sum_{N_{1}=0}^{N_{1}=\infty}\sum_{N_{2}=0}^{N_{2}=\infty}...\sum_{N_{k}=0}^{N_{k}=
\infty}\mathcal{P}\left(N_{1},N_{2},...,N_{k}\right)
\int{dx_{1}^{1}P_{1}(x_{1}^{1})}...\int{dx_{N_{1}}^{1}P_{1}(x_{N_{1}}^{1})}\nonumber\\
&\times\int{dx_{1}^{2}P_{2}(x_{1}^{2})}...
\int{dx_{N_{2}}^{2}P_{2}(x_{N_{2}}^{2})}\times...\times\int{dx_{1}^{k}P_{k}(x_{1}^{k})}...
\int{dx_{N_{k}}^{k}P_{k}(x_{N_{k}}^{k})} \nonumber
\\
& \times  \left[w_{p}(x_{1}^{1})
+...+w_{p}(x_{N_{1}}^{1})+w_{p}(x_{1}^{2})+...+w_{p}(x_{N_{2}}^{2})
+w_{p}(x_{1}^{k})+...+w_{p}(x_{N_{k}}^{k})\right]\nonumber\\
&\times \left[w_{q}(x_{1}^{1})+...+w_{q}(x_{N_{1}}^{1})+w_{q}(x_{1}^{2})+...
+w_{q}(x_{N_{2}}^{2})+w_{q}(x_{1}^{k})+...+w_{q}(x_{N_{k}}^{k})\right]\nonumber\\
&\times
\left[w_{r}(x_{1}^{1})+...+w_{r}(x_{N_{1}}^{1})+w_{r}(x_{1}^{2})+...
+w_{r}(x_{N_{2}}^{2})+w_{r}(x_{1}^{k})+...+w_{r}(x_{N_{k}}^{k})\right]\nonumber\\
%
%
%\eq{
&=\sum_{i=1}^{k}\left<N_{i}^{3}\right>W^{pqr}_{iii}
%\nonumber\\
%
+  \sum_{1\le i < l \le
k}\left<N_{i}^{2}N_{l}\right>\left(W^{pqr}_{iil} + W^{pqr}_{ili} +
W^{pqr}_{lii}\right)
\nonumber \\
& +  \sum_{1\le i < l \le k}\left<N_{i}N_{l}^{2}\right>
\left(W^{pqr}_{lli} + W^{pqr}_{lil} + W^{pqr}_{ill}\right)\nonumber\\
&+ \sum_{1\le i < l < m \le
k}\left<N_{i}N_{l}N_{m}\right>\left(W^{pqr}_{ilm}
+ W^{pqr}_{lmi} + W^{pqr}_{mil} + W^{pqr}_{iml}
+ W^{pqr}_{lim} + W^{pqr}_{mli}\right)\nonumber\\
&+\sum_{i=1}^{k}\left<N_{i}^{2}\right>\left(\overline{w_{pq,i}}\cdot\overline{w_{r,i}}
+ \overline{w_{pr,i}}\cdot\overline{w_{q,i}} +
\overline{w_{qr,i}}\cdot\overline{w_{p,i}} -3\overline{w_{p,i}}\cdot\overline{w_{q,i}}\cdot\overline{w_{r,i}}\right)\nonumber\\
%
%} \eq{
%
&+\sum_{1\le i < l \le
k}\left<N_{i}N_{l}\right>\left(W^{(pq)r}_{il} + W^{(pq)r}_{li} +
W^{(pr)q}_{il} + W^{(pr)q}_{li} + W^{(qr)p}_{il} +
W^{(qr)p}_{li}\right)~\nonumber \\
&+ \sum_{i=1}^{k}\left<N_{i}\right>\left(\overline{w_{pqr,i}} +
2W^{pqr}_{iii} - \overline{w_{pq,i}}\cdot\overline{w_{r,i}} -
\overline{w_{pr,i}}\cdot\overline{w_{q,i}} -
\overline{w_{qr,i}}\cdot\overline{w_{p,i}}\right)~,
}
%
%\begin{multline}
%
\eq{
& \left<W_{p}^{2}W_{q}\right>
=\sum_{N_{1}=0}^{N_{1}=\infty}\sum_{N_{2}
=0}^{N_{2}=\infty}...\sum_{N_{k}=0}^{N_{k}
=\infty}\mathcal{P}\left(N_{1},N_{2},...,N_{k}\right)
\int{dx_{1}^{1}P_{1}(x_{1}^{1})}...\int{dx_{N_{1}}^{1}P_{1}(x_{N_{1}}^{1})}\nonumber\\
&\times\int{dx_{1}^{2}P_{2}(x_{1}^{2})}...
\int{dx_{N_{2}}^{2}P_{2}(x_{N_{2}}^{2})}\times...
\times\int{dx_{1}^{k}P_{k}(x_{1}^{k})}...\nonumber \\
&\times
\int{dx_{N_{k}}^{k}P_{k}(x_{N_{k}}^{k})}\left[w_{p}(x_{1}^{1})+...+w_{p}(x_{N_{1}}^{1})
+w_{p}(x_{1}^{2})+...+w_{p}(x_{N_{2}}^{2})
+w_{p}(x_{1}^{k})+...+w_{p}(x_{N_{k}}^{k})\right]^{2}\nonumber\\
&\times \left[w_{q}(x_{1}^{1})+...+w_{q}(x_{N_{1}}^{1})
+w_{q}(x_{1}^{2})+...+w_{q}(x_{N_{2}}^{2})+w_{q}(x_{1}^{k})+...
+w_{q}(x_{N_{k}}^{k})\right]\nonumber \\
&=
\sum_{i=1}^{k}\left<N_{i}^{3}\right>\overline{w_{p,i}}^{2}\cdot\overline{w_{q,i}}
 +  \sum_{1\le i < l \le
k}\left<N_{i}^{2}N_{l}\right>\left(\overline{w_{p,i}}^{2}\cdot\overline{w_{q,l}}
+  2W^{ppq}_{ili}\right)
+  \sum_{1\le i < l \le k}\left<N_{i}N_{l}^{2}\right>
\left(\overline{w_{p,l}}^{2}\cdot\overline{w_{q,i}}  +  2W^{ppq}_{ill}\right)\nonumber \\
&+2\sum_{1\le i < l < m \le k}\left<N_{i}N_{l}N_{m}\right>
\left(W^{ppq}_{ilm} + W^{ppq}_{iml} + W^{ppq}_{lmi}\right)%\\
+\sum_{i=1}^{k}\left<N_{i}^{2}\right>\left(\overline{w_{p,i}^{2}}\cdot\overline{w_{q,i}}
- 3\overline{w_{p,i}}^{2}\cdot\overline{w_{q,i}} +
2\overline{w_{qp,i}}\cdot\overline{w_{p,i}}\right)\nonumber\\
&+ \sum_{1\le i < l \le
k}\left<N_{i}N_{l}\right>\left(2W^{(pq)p}_{il} + 2W^{(pq)p}_{li} +
\overline{w_{p,i}^{2}}\cdot\overline{w_{q,l}} +
\overline{w_{p,l}^{2}}\cdot\overline{w_{q,i}} -
\overline{w_{p,i}}^{2}\cdot\overline{w_{q,l}}
-\overline{w_{p,l}}^{2}\cdot\overline{w_{q,i}}\right)\nonumber\\
%\
%}
%
%\eq{
%
 &+
\sum_{i=1}^{k}\left<N_{i}\right>\left(2\overline{w_{p,i}}^{2}\cdot\overline{w_{q,i}}
+ \overline{w_{pq,i}^{2}} -
\overline{w_{p,i}^{2}}\cdot\overline{w_{q,i}} -
2\overline{w_{pq,i}}\cdot\overline{w_{p,i}}\right)~.\label{Wp2Wq}
}
By exchanging in the last equation $p$ by $q$ one gets  a similar
expression for the $\left<W_{p}W_{q}^{2}\right>$. In
Eqs.~(\ref{Wp3}-\ref{Wp2Wq}) the following notations are used:
%
%\begin{equation*}
\eq{\label{1}
& W^{abc}_{\alpha\beta\gamma} =
\overline{w_{a,\alpha}}\cdot\overline{w_{b,\beta}}\cdot\overline{w_{c,\gamma}}~,~~~~
%\end{equation*}
%
%\begin{equation*}
 W^{(ab)c}_{\beta\gamma} =
\overline{w_{ab,\beta}}\cdot\overline{w_{c,\gamma}} -
\overline{w_{a,\beta}}\cdot\overline{w_{b,\beta}}\cdot\overline{w_{c,\gamma}}~,\\
%\end{equation*}
%\begin{equation*}
&
\overline{w_{i,j}^{n}}=\frac{1}{\left<N_{j}\right>}\int{w_{i}^{n}(x)\rho_{j}(x)dx}~,~~~~
%\end{equation*}
%
%\begin{equation*}
\overline{w_{pq,i}^{n}}=\frac{1}{\left<N_{i}\right>}\int{w_{p}^{n}(x)w_{q}(x)\rho_{i}(x)dx}~,\label{2}\\
%\end{equation*}
%\begin{equation*}
&
\overline{w_{pqr,i}}=\frac{1}{\left<N_{i}\right>}\int{w_{p}(x)w_{q}(x)w_{r}(x)\rho_{i}(x)dx}~.\label{3}
%\end{equation*}
}

As it is evident from Eqs.~(\ref{Wp3}-\ref{Wp2Wq}) the third
moments of the $W$-quantities $\left<W_{p}^{3}\right>$,
$\left<W_{p}^{2}W_{q}\right>$, $\left<W_{q}^{2}W_{p}\right>$, and
$\left<W_{p}W_{q}W_{r}\right>$ are just linear combinations of
the unknown third moments $\left<N_{p}^{3}\right>$,
$\left<N_{p}^{2}N_{q}\right>$, $\left<N_{q}^{2}N_{p}\right>$, and
$\left<N_{p}N_{q}N_{r}\right>$ of the multiplicity distributions. In Eqs.~(\ref{Wp3}-\ref{Wp2Wq})
the first moments $\langle N_j\rangle$ and second moments $\langle
N_j^2\rangle$, $\langle N_pN_q\rangle$ are assumed to have already been determined, as it is
discussed in Section~\ref{second_moments}. The coefficients
(\ref{1}-\ref{3}) entering into the right-hand-side of
Eqs.~(\ref{Wp3}-\ref{Wp2Wq}) are also calculable, since they are
expressed in terms of experimentally measured
$\rho_j(x)$-functions.  The problem is then reduced to solving the
system of linear equations for the third moments of the
multiplicity distribution in terms of experimentally measured
third moments of $W$-quantities. This approach gives a unique
answer, because by construction the number of equations is equal
to the number of unknown third moments.

The method can be used to calculate any moments of the joint
multiplicity distribution by deriving the corresponding set of
linear equations.

%\section{Error propagation}
\section{First moments reexamined}\label{first_revisit}
As mentioned in sections~\ref{def} and~\ref{second_moments}
the first moments of multiplicity distributions can be directly
obtained by integrating the corresponding $x$ distributions,
$\rho_j(x)$. Here we consider the case when the shape of the 
$\rho_j(x)$ functions are known, whereas their normalizations are
arbitrary. This may be important for the analysis of hadron
production in a narrow acceptance region. One then uses the data
from the wider kinematic region to enlarge statistics and thus to 
improve the  quality of the $\rho_j(x)$ functions.
We denote these arbitrary normalizations by $A_{j}$, i.e,
 \eq{ \label{norm-rho-i-A} \int dx \,\rho_j (x) = \langle A_j
\rangle~.
}
It should be emphasized that all identities defined in
Eq.~(\ref{wi}) should be recalculated using the distributions
defined in Eq.~(\ref{norm-rho-i-A}).

Below we demonstrate that the identity method can also be applied
to obtain the hadron multiplicities with these arbitrarily
normalized mass distributions. The idea is to extract the unknown
first moments of multiplicity distributions from first moments of
$W$-quantities.
Indeed, one can write:
%
%\begin{multline}
\eq{\label{Wp} &
\left<W_{p}\right>=\sum_{N_{1}=0}^{N_{1}=\infty}\sum_{N_{2}=0}^{N_{2}
 =\infty}...\sum_{N_{k}=0}^{N_{k}=\infty}\mathcal{P}\left(N_{1},N_{2},...,N_{k}\right)
 \int{dx_{1}^{1}P_{1}(x_{1}^{1})}...\int{dx_{N_{1}}^{1}P_{1}(x_{N_{1}}^{1})}\nonumber \\
&\times\int{dx_{1}^{2}P_{2}(x_{1}^{2})}...\int{dx_{N_{2}}^{2}P_{2}(x_{N_{2}}^{2})}\times...
\times\int{dx_{1}^{k}P_{k}(x_{1}^{k})}...\int{dx_{N_k}^{k}P_{k}(x_{N_k}^{k})}\nonumber \\
&\times \left[w_{p}(x_{1}^{1})+...+w_{p}(x_{N_{1}}^{1})+w_{p}(x_{1}^{2})+...+w_{p}(x_{N_{2}}^{2})
+w_{p}(x_{1}^{k})+...+w_{p}(x_{N_{k}}^{k})\right]\nonumber \\
&= \sum_{i=1}^{k}\left<N_{i}\right>\overline{w_{p,i}}~,
%\end{multline}
}
where the
mass probability distribution functions are now defined as
$P_{i}(x)\equiv\rho_{i}(x)/\left<A_{i}\right>$, and
%\begin{equation*}
\eq{\label{wpi}
\overline{w_{p,i}}=\frac{1}{\left<A_{i}\right>}\int{w_{p}(x)\rho_{i}(x)dx}~.
%\end{equation*}
}
The equation above can be written in a matrix form:
 \eq{\label{wpimatrix}
\left( \begin{array}{c}
      \left<W_1\right>   \\
      \left<W_2\right>  \\
      ...\\
      \left<W_k\right>
      \end{array} \right)=
\left(\begin{array}{cccc} \overline{w_{1,1}} & \overline{w_{1,2}} & ... & \overline{w_{1,k}}   \\
      \overline{w_{2,1}} & \overline{w_{2,2}} & ... & \overline{w_{2,k}} \\
      ...& ... & ... & ...\\
      \overline{w_{k,1}} & \overline{w_{k,2}} & ... & \overline{w_{k,k}}
      \end{array} \right) \left( \begin{array}{c}
      \left<N_1\right>   \\
      \left<N_2\right>  \\
      ...\\
      \left<N_k\right>
      \end{array} \right) ~.
}
The only unknowns in Eq.~(\ref{wpimatrix}) are the first moments
$\left<N_{j}\right>$ which can easily be obtained by matrix
inversion.

Note that when $\left<A_{i}\right>$ = $\left<N_{i}\right>$ the right-hand-side
of Eq.~(\ref{Wp}) yields:
%
%
%\begin{equation*}
\eq{\label{Ni}
\sum_{i=1}^{k}\left<N_{i}\right>\overline{w_{p,i}}=
\sum_{i=1}^k\int{\frac{\rho_{p}\cdot\rho_{i}}{\rho}dx} =
%\int{\frac{\rho_{p}*\sum\rho_{i}}{\rho}dm} =
\int{\rho_{p}dx} = \left<N_{p}\right>~.
}
%\end{equation*},
%
Therefore, in this special case the first moments of the
multiplicity distributions are identical to the first moments of
$W$-quantities,
%
%\begin{equation*}
%\eq{ \label{Np}
%
$ \left<W_{p}\right>=\left<N_{p}\right>$~,
%\end{equation*}
%}
%
similar to the case of CI, cf. Eq.~(\ref{CI}).

\section{Test on simulated data \label{sim}}

%\begin{figure}[htb]
 %{\includegraphics[width=0.8\linewidth,clip=true]
   %   {fig1N}
%} \caption{(Color online) Inclusive distribution function of the
%particle identification variable $x$ for proton (red), kaon (blue)
%and pion(green) used as an input to our simulations.} \label{fig1}
%\end{figure}

\begin{figure}[htb]
 {\includegraphics[width=0.8\linewidth,clip=true]
      {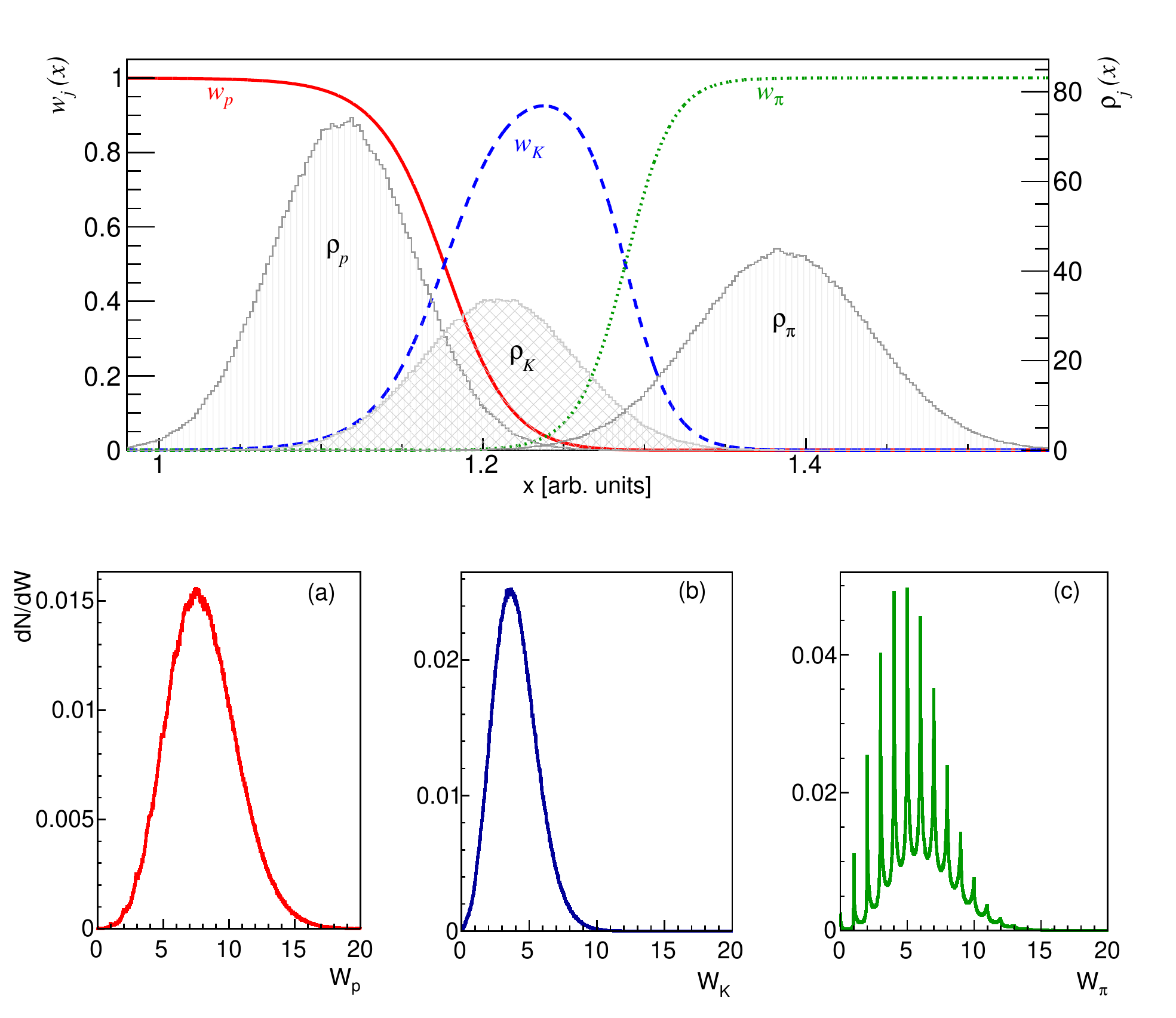}
} \caption{(Color online) Upper panel: The inclusive distribution $\rho_{j}(x)$
of the particle identification variable $x$ for protons, kaons
and pions, used as an input to the
simulation, are depicted with shaded histograms. The
distributions of identities $w_{j}(x)$ for protons, kaons and pions
are presented with red solid, blue dashed and green dotted lines respectively.
Lower panel: The distribution of event variables $W_{j}$ with $j$
standing for (a) protons, (b) kaons and (c) pions.} \label{fig1}
\end{figure}

In this section we apply formulae (7-9) to simulated data. As already mentioned in section~\ref{def}
we assume that particle identification is achieved by
measuring a quantity $x$. In each event we generate three
different particle species\footnote{Note that the method does not
have any limitation on the number of particle species.} called
proton (p), kaon (K), and pion ($\pi$). Furthermore, the
multiplicity distributions of these particles are simulated
according to the Poisson distribution with different mean values. Consequently,
the input information to our simulation is an inclusive (i.e.,
summed over all events and all tracks) distribution of the PID
variable $x$ for each particle (see shaded histograms in upper panel of Figure~\ref{fig1}) and mean
multiplicities of particles. In our example these multiplicities
are 8, 6 and 4 for protons, pions and kaons respectively.  The simulation process comprises the 
following: (i) in each event we randomly generate from the
Poisson distribution particle multiplicities using their mean
values; (ii) we generate the quantity $x$ for each particle $j$ in
an event from the inclusive distribution $\rho_j(x)$ of $x$
quantity for the corresponding particle. Using this simulated data
we reconstruct all second and third moments of the multiplicity
distributions. For each measurement $x$ in an event we calculate
identity variables $w_{j}(x)$ according to Eq.~(\ref{wi}), where
$j$ stands for the particle type. Next, for each particle we
determine the event variable $W_{j}$ (see Eq.~(\ref{Wj-def})).
Figure~\ref{fig1} (upper panel) shows the distributions of the
identity variables $w_{j}$ for different particle species. In the
lower panel of Fig.~\ref{fig1} distributions of the $W_{j}$
quantities are plotted. As seen from Fig.~\ref{fig1} the
$W_{\pi}$ distribution exhibits evident structures. This is not
surprising as by construction the latter is nothing else but the
original multiplicity distribution smeared because of non-ideal
particle identification, i.e., because of  the overlapping
$\rho_j(x)$ distributions of different particles. These structures
are better seen for $W_{\pi}$ because the $x$ distribution of
the pions is well separated from those of protons and kaons as
illustrated by the shaded histograms in the upper panel of
Fig.~\ref{fig1}.

Finally using Eqs.~(7-9) we reconstruct all third moments of the
particles from the corresponding moments of the W quantities. We
would like to emphasize that all second moments needed for
reconstruction of the third moments are obtained using the formulae of
Ref.~\cite{Go2011}. The results are summarized in Table I, where
the Poisson values are calculated using the moment generating
function of the Poisson distribution, $M(t) = e^{-<N>}e^{<N>
e^{t}}$:
\eq{\label{p2}<N^{2}> = M''(t=0) = <N>^{2}+<N>,}
\eq{ \label{p3}
<N^{3}> = M'''(t=0) = <N>^{3}+3<N^{2}>+<N>.}

\begin{table*}
%\begin{center}
\begin{tabular}{|c|c|c|c|}
\hline \hline  Moments & Reconstructed with the identity method & Generated values & Poisson values \\ [0.5ex] 
\hline  
$\left< p^2 \right> $ & 72.0429 $\pm$  0.0122 & 72.0452 &72 \\  [0.5ex] 
\hline  
$\left< \pi^2 \right> $ & 42.0033 $\pm$  0.0073 & 42.0011 & 42 \\  [0.5ex] 
\hline  
$\left< K^2 \right> $ & 20.0066 $\pm$  0.0052 & 20.0057 & 20 \\  [0.5ex] 
\hline  
$\left< p\pi\right> $ & 48.0178 $\pm$  0.007 & 48.0184 &48 \\  [0.5ex] 
\hline  
$\left<p K \right> $ & 32.0147 $\pm$  0.0057 & 32.0154 & 32 \\  [0.5ex]
 \hline
 $\left< \pi K \right> $ & 24.0026 $\pm$  0.0033 & 24.002 & 24 \\  [0.5ex]
\hline 
\hline 
\hline  
$\left< p^3 \right> $ & 712.6391 $\pm$ 0.1925 & 712.6718 & 712 \\  [0.5ex] 
\hline  
$\left< \pi^3 \right> $ & 330.0547 $\pm$  0.0878 & 330.0316 & 330 \\  [0.5ex] 
\hline  
$\left< K^3\right> $ & 116.045 $\pm$  0.0501 & 116.0372 &116 \\  [0.5ex] 
\hline
$\left< p^2\pi \right> $ & 432.307 $\pm$  0.0954 & 432.3185 & 432 \\ [0.5ex] 
\hline  
$\left< p^2 K\right> $ & 288.1968 $\pm$  0.0710 & 288.2081 & 288 \\  [0.5ex] 
\hline  
$\left< \pi^2 K \right> $ & 168.0083 $\pm$ 0.033 & 167.9971 & 168 \\  [0.5ex] 
\hline  
$\left< \pi^2 p\right> $ & 336.155 $\pm$  0.0786 & 336.1525 & 336 \\  [0.5ex] 
\hline  
$\left< K^2 p\right> $ & 160.0986 $\pm$  0.0518 & 160.0875 & 160 \\  [0.5ex] 
\hline
$\left< K^2 \pi \right> $ & 120.0275 $\pm$  0.0320 & 120.0174 & 120 \\ [0.5ex] 
\hline  $\left< p \pi K \right> $ & 192.0765 $\pm$  0.0379 & 192.0845 & 192 \\  [0.5ex] 
\hline
\hline
\end{tabular}
%\end{center}
\caption{Results from the identity method for the
15$\cdot$10$^{6}$ simulated events   with mean multiplicities
corresponding to $<p>= 8, <\pi> = 6$ and $<K> = 4$. All second
moments (upper part) listed in the table are reconstructed according to Ref.~\cite{Go2011}, all third moments
     (lower part) according to the formulae of Section~\ref{higher}. The
     statistical errors were obtained from the Monte Carlo method
     using Eqs.(23-25). Moments directly calculated from the
     simulated multiplicity distributions are shown in the third
     column and the Poisson values obtained from Eqs.(18-22) are
     listed in the last column.}
\label{table1}
\end{table*}

\begin{table*}
%\begin{center}
\begin{tabular}{|c|c|c|c|}
\hline \hline  Moments & Reconstructed with the identity method & Generated values & Poisson values \\ [0.5ex] 
\hline  
$\left< p^2 \right> $ & 72.0216 $\pm$  0.0172 & 72.0198 &72 \\  [0.5ex] 
\hline  
$\left< \pi^2 \right> $ & 42.0108 $\pm$  0.0066 & 42.0054 & 42 \\  [0.5ex] 
\hline  
$\left< K^2 \right> $ & 72.0177 $\pm$  0.0144 & 72.0198 & 72 \\  [0.5ex] 
\hline  
$\left< p\pi\right> $ & 48.0154 $\pm$  0.0082 & 48.0096 &48 \\  [0.5ex] 
\hline  
$\left<p K \right> $ & 72.0166 $\pm$  0.0143 & 72.0198 & 72 \\  [0.5ex]
 \hline
 $\left< \pi K \right> $ & 48.0074 $\pm$  0.0072 & 48.0096 & 48 \\  [0.5ex]
\hline 
\hline 
\hline  
$\left< p^3 \right> $ & 712.3745 $\pm$ 0.2572 & 712.3739 & 712 \\  [0.5ex] 
\hline  
$\left< \pi^3 \right> $ & 330.0893 $\pm$  0.0806 & 330.0229 & 330 \\  [0.5ex] 
\hline  
$\left< K^3\right> $ & 712.3043 $\pm$  0.2167 & 712.3739 & 712 \\  [0.5ex] 
\hline
$\left< p^2\pi \right> $ & 432.2096 $\pm$  0.1246 & 432.1387 & 432 \\ [0.5ex] 
\hline  
$\left< p^2 K\right> $ & 712.3465 $\pm$  0.2249 & 712.3739 & 712 \\  [0.5ex] 
\hline  
$\left< \pi^2 K \right> $ & 336.0526 $\pm$ 0.0744 & 336.0604 & 336 \\  [0.5ex] 
\hline  
$\left< \pi^2 p\right> $ & 336.1102 $\pm$  0.0838 & 336.0604 & 336 \\  [0.5ex] 
\hline  
$\left< K^2 p\right> $ & 712.3191 $\pm$  0.2079 & 712.3739 & 712 \\  [0.5ex] 
\hline
$\left< K^2 \pi \right> $ & 432.0899 $\pm$  0.1054 & 432.1387 & 432 \\ [0.5ex] 
\hline  $\left< p \pi K \right> $ & 432.1456 $\pm$  0.1079 & 432.1387 & 432 \\  [0.5ex] 
\hline
\hline
\end{tabular}
%\end{center}
\caption{Results from the identity method for the
15$\cdot$10$^{6}$ simulated events   with mean multiplicities
corresponding to $<p>= 8, <\pi> = 6$. In each event the number of kaons is taken to be equal to the number of protons in order to probe the method in presence of correlations. 
All second moments (upper part) listed in the table are reconstructed according to Ref.~\cite{Go2011}, all third moments (lower part) according to the formulae of 
Section~\ref{higher}. The statistical errors were obtained from the Monte Carlo method using Eqs.(23-25). Moments directly calculated from the
simulated multiplicity distributions are shown in the third column and the Poisson values are listed in the last column. 
Note that for protons and kaons Eqs.(20-22) are not satisfied anymore due to the introduced correlations.}
\label{table2}
\end{table*}

In case of vanishing correlations between particles (as assumed in our simulation) one finds:
\eq{\label{p4} <N_{1}^{2}N_{2}> =  (<N_{1}>^{2}+<N_{1}>)<N_{2}>,}
\eq{\label{p5}<N_{1}N_{2}> =  <N_{1}><N_{2}>,}
\eq{\label{p6}<N_{1}N_{2}N_{3}> =  <N_{1}><N_{2}><N_{3}>.}
The statistical errors of the reconstructed moments of the
multiplicity distributions result from the errors on the
parameters of the fitted distributions $\rho_{j}(x)$ of the identification observable $x$ and
the errors of the $W_{j}$ quantities\footnote{Contributions to the uncertainty of the reconstructed
 moments due to the assumptions on the functional representation
 of $\rho_{j}(x)$ depend on the features of the experiment and their
 discussion is beyond the scope of this paper.}. Typically these two sources
of errors are correlated. Moreover, one can experimentally
study fluctuation signals built up from several moments of
the multiplicity distributions. The standard error propagation is complicated and inconvenient.
We therefore follow the Monte Carlo error calculation
approach.  We first randomly subdivide the simulated data into n
subsamples and for each subsample reconstruct the moments $M_{n}$
listed in Table I. In the second step we calculate the statistical
error of each moment $M$ as:
\eq{ \label{stat-1}
 \sigma_{<\textit{M}>}=\frac{\sigma}{\sqrt{n}},
}
where
\eq{ \label{stat-2}
 <\textit{M}>=\frac{1}{n}\sum{\textit{M}_n} ,
}
and
\eq{ \label{stat-3}
 \sigma=\sqrt{\frac{\sum\left(\textit{M}_i-<\textit{M}>\right)^2}{n-1}}.
}

\begin{figure}[htb]
 {\includegraphics[width=0.7\linewidth,clip=true]
      {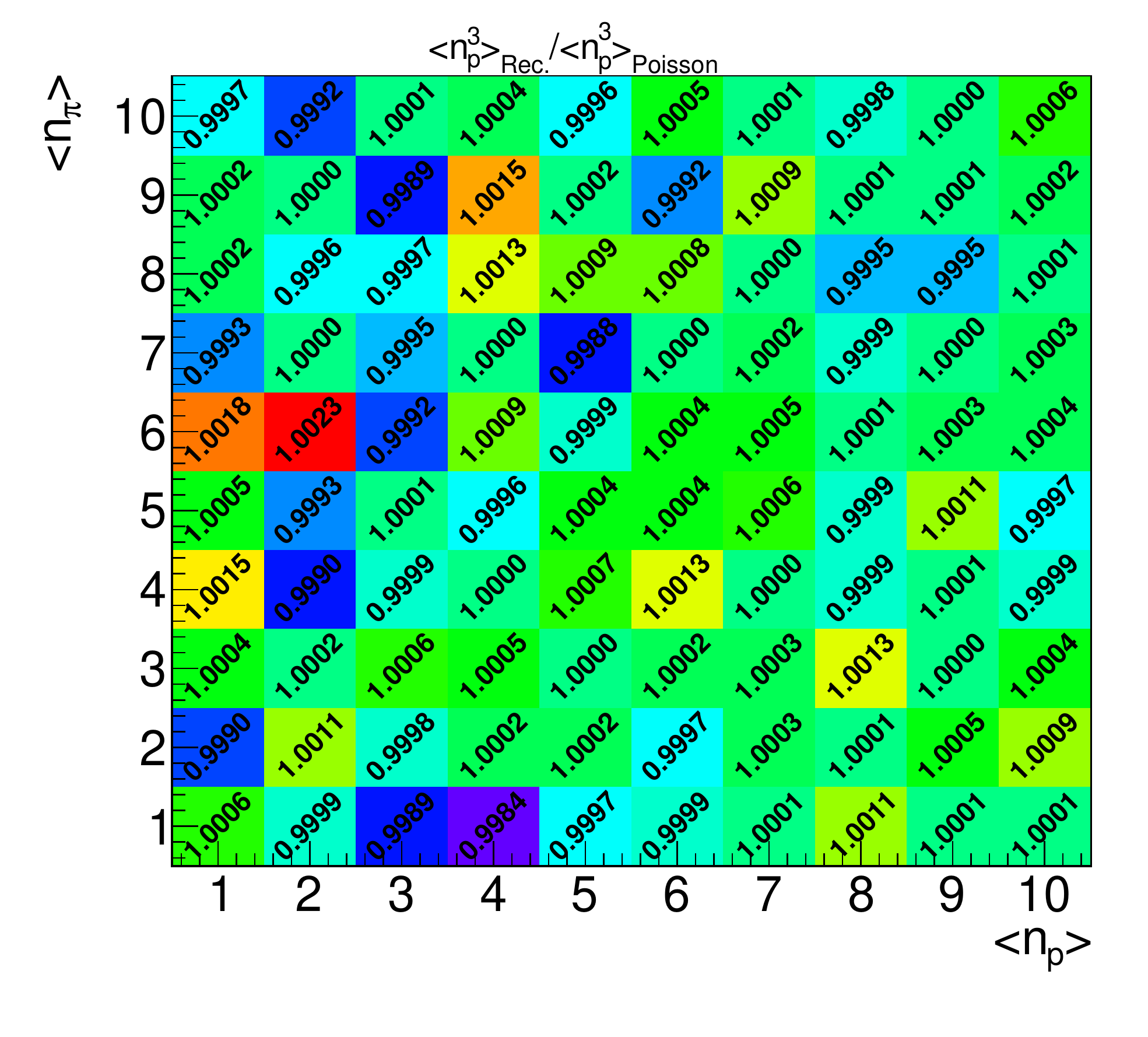}
} \caption{(Color online) Result of 100 simulations, each with 6$\cdot$10$^{6}$ events, with different
mean multiplicities of protons and pions while the 
multiplicity of kaons in each event  is taken to  be equal to the number of protons. In each box the
ratio of the reconstructed third moment $<n_{p}^{3}>$ to the
Poisson expectation (see Eq.~(\ref{p3})) is presented.
} \label{fig2}
\end{figure}

The reconstructed moments, using the identity method, presented in Table~\ref{table1} are in agreement with the generated moments within statistical uncertainties. 
Next, in order to demonstrate that the method works also in case of strong correlations, we introduced an artificial correlation between protons and kaons. In each event we simulate the number of the protons and pions from the corresponding Poisson distributions as described above. However, the number of the kaons in each event is taken to be equal to the number of protons. The reconstructed moments from this simulation for 15$\cdot$10$^6$ events with the mean multiplicities corresponding to 8 and 6 for the protons and pions respectively are presented in Table~\ref{table2}. Note that the mean multiplicity of kaons is the same as for protons as described above. As evident from the Table~\ref{table2} the reconstructed moments are again in reasonable agreement with the generated moments. However, for protons and kaons Eqs.(20-22) are not satisfied anymore due to the introduced correlations.

In Fig.~\ref{fig2} we demonstrate the result of 100 simulations, with 6$\cdot$10$^{6}$ events in each,
with different mean multiplicities of protons and pions. The number of the kaons in each event is taken to be equal to the number of protons, in order to induce a correlation between particles (see the text above). In each box the ratio of the reconstructed third moment
$<n_{p}^{3}>$ of the proton multiplicity distribution to the Poisson expectation calculated using Eq.~(\ref{p3}) is presented. One finds a high
precision of the identity method independent of the values of
average multiplicities.

\section{Summary}\label{sum}

In summary, we extended the identity method proposed in
Refs.~\cite{Ga2011,Go2011} to third moments of unknown
multiplicity distributions. We introduce the event  quantities
$W_j$ depending on the single-particle identity variables $w_j$
according to Eq.~(\ref{Wj-def}).  Any moments of these quantities
$\langle W_p\ldots W_q\rangle$ can be measured experimentally.
The problem  of finding the second moments $\langle N_j^2\rangle$
and $\langle N_pN_q\rangle$ was solved in Ref.~\cite{Go2011}.
The third moments of $W$-quantities are found to be linear
combinations of the unknown third moments
$\left<N_{p}^{3}\right>$, $\left<N_{p}^{2}N_{q}\right>$ ,
$\left<N_{q}^{2}N_{p}\right>$ and $\left<N_{p}N_{q}N_{r}\right>$.
The first moments $\langle N_j\rangle$ and second moments $\langle
N_j^2\rangle$, $\langle N_pN_q\rangle$ entering into these
relations are already known. We are confident that the procedure
can be used for obtaining any higher moments of the multiplicity
distributions by deriving the corresponding set of linear
equations.
We thus construct an iterative procedure in which all moments of
the multiplicity distribution of the $n^{th}$ order are calculated
from measured $n^{th}$ order moments of the $W$-quantities.
% all moments of the multiplicity distribution up
%to the order of $n-1$ are found at the previous steps of this
%procedure.
By doing so, we get a unique solution, as the number of linear
equations in each step is always equal to the number of the
unknown moments.
We presented an explicit test of the method using simulated events including the case of strong correlations between particles. 
%Moreover the identity method for the second moments has successfully been applied to the experimental data.
%
Note that the higher moments (e.g., third and fourth moments of
pion and proton multiplicity distributions \cite{Step,Step1}) can
reveal the effects of the QCD critical point with
higher sensitivity in the event-by-event multiplicity fluctuations
in nucleus-nucleus collisions

%
%the signals of the QCD critical point from the event-by-event
%fluctuations in nucleus-nucleus collisions.

\begin{acknowledgments} We would like to thank
Marek Ga\'zdzicki, Peter Seyboth and Herbert Str\"{o}bele for fruitful
discussions and comments.  The work of M.~I.~G. was supported by
Humboldt Foundation and by the Program of Fundamental Research of
the Department of Physics and Astronomy of NAS, Ukraine. A.~R.
gratefully acknowledges the support by the German Research
Foundation (DFG grant GA 1480/2.1).
\end{acknowledgments}

%\bibliographystyle{unsrt}

%\bibliography{rustamov}

\end{document}